\documentclass[preliminary,copyright,creativecommons]{eptcs}
\providecommand{\event}{G.Ciobanu, M.Koutny (Eds.): Membrane Computing and\\
Biologically Inspired Process Calculi (MeCBIC 2010)\\
EPTCS Preliminary Proceedings, 2010.} 
\usepackage{graphicx}
\usepackage{array}
\usepackage{color}
\usepackage{wrapfig}
\usepackage{multirow}
\input {pdfcolor.tex}

\title{Computational Modeling for the Activation Cycle of G-proteins
    by G-protein-coupled Receptors\\ {\small In memory of Robin and
      Lucy Milner}}
\author{Yifei Bao\qquad \qquad Adriana B.\  Compagnoni \thanks{We are
    grateful to Andrew Phillips of Microsoft Research for his help
    with understanding stochastic rates translation and improving our Pi-calculus
    model, and for his insightful comments on the classification of 
    modeling methods, Vishakha Sharma and Thomas Cattabiani of Stevens Institute of Technology for
    critical reading of an earlier version of this manuscript, and Amanda DiGuilio of Stevens
    Institute of Technology for her help with understanding the
    biology. We are also grateful to the anonymous reviewers for
    helpful commenst and valuable suggestions for future work.}
\institute{Department of Computer Science\\
Stevens Institute of Technology\\
 Hoboken, USA}
\email{\quad ybao@stevens.edu  \quad\qquad abc@cs.stevens.edu}
\and
Joseph S.\  Glavy\qquad\qquad Tommy E.\  White
\institute{Department of Chemical Biology and Biomedical Engineering\\
Stevens Institute of Technology\\
 Hoboken, USA}
\email{\quad  jglavy@stevens.edu \quad\qquad twhite1@stevens.edu}\\
}
\def\titlerunning{Computational Modeling for the Activation Cycle of G-proteins
    by G-protein-coupled Receptors}
\def\authorrunning{Y.\ Bao, A.\ B.\ Compagnoni, J.\ S.\ Glavy \& T.\ E.\ White}
\begin{document}
\maketitle

\begin{abstract} 
  In this paper, we survey five different computational modeling
  methods. For comparison, we use the activation cycle of G-proteins
  that regulate cellular signaling events downstream of
  G-protein-coupled receptors (GPCRs) as a driving example. Starting
  from an existing Ordinary Differential Equations (ODEs) model, we
  implement the G-protein cycle in the stochastic Pi-calculus using
  SPiM, as Petri-nets using Cell Illustrator, in the Kappa Language
  using Cellucidate, and in Bio-PEPA using the Bio-PEPA eclipse plug
  in. We also provide a high-level notation to abstract away from
  communication primitives that may be unfamiliar to the average
  biologist, and we show how to translate high-level programs into
  stochastic Pi-calculus processes and chemical reactions.
\end{abstract}

\section{Introduction}\label{intro}

Traditionally, biologists have used ordinary differential equations
(ODEs) to model biological processes and simulate the evolution of
species over time. However, the past decade has seen the emergence of
a family of formalisms dedicated to modeling biological behavior
(kinetics). Stemming from formal languages designed to capture
concurrent computation and communicating processes, these new
formalisms range from graphical formalisms to algebraic languages.

In this paper we survey five different computational modeling formalisms,
and we show how to simulate the activation cycle of G-proteins by
G-protein coupled receptors in each of them.

\subsection{The G-protein cycle}
G-protein-coupled receptors (GPCRs) are
seven-pass transmembrane receptors \cite{Pierce02} that mediate
extracellular signaling molecules and intracellular signal
transduction pathways via trimeric GTP binding proteins (G
proteins) \cite{Lousi06}. It is estimated that over half of all
marketed pharmaceuticals target GPCRs \cite{Howard01}.

\begin{figure}[!htb]
\begin{center}
\includegraphics[scale=0.11]{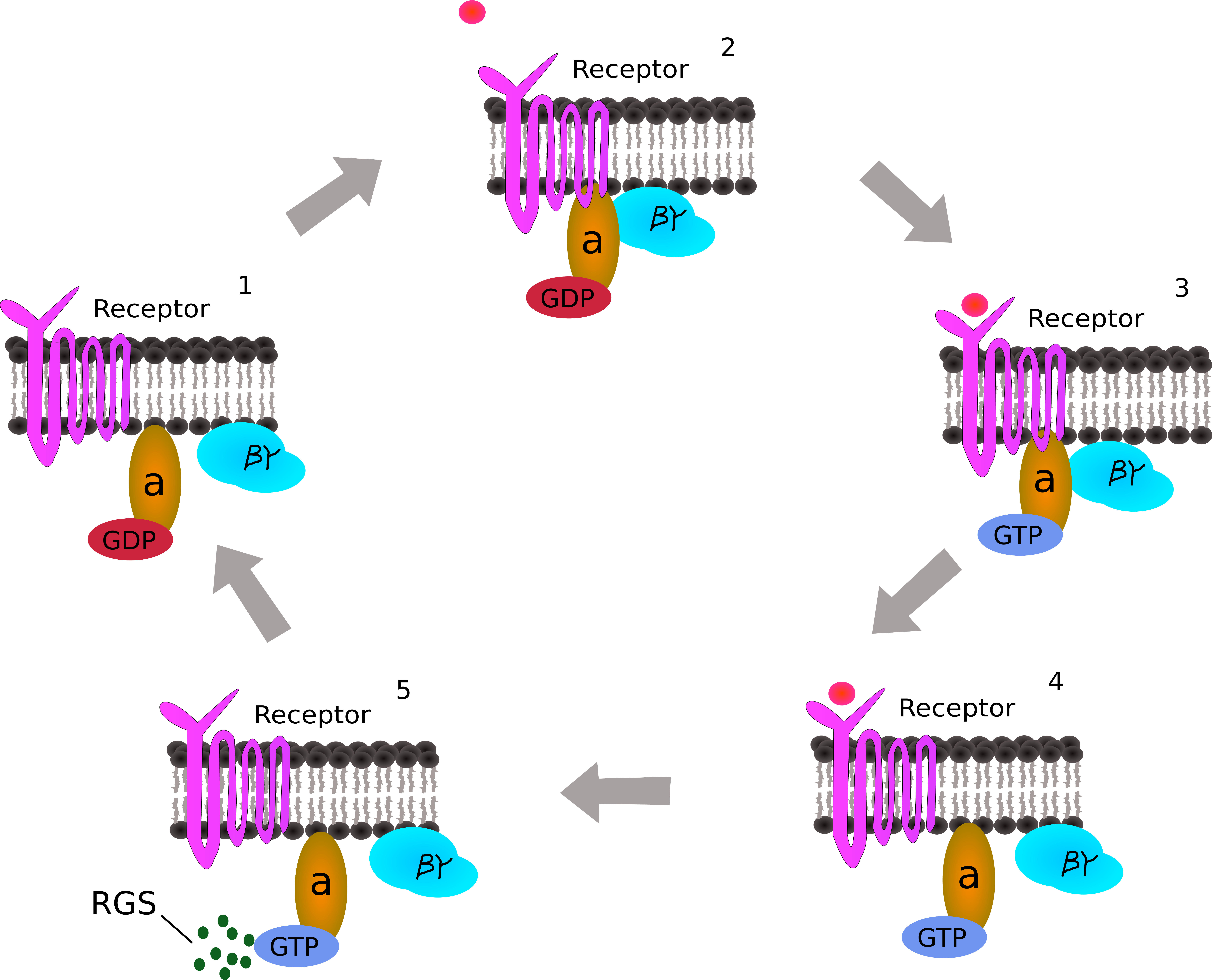}
\end{center}
\caption{\footnotesize{Activation cycle of G-proteins by
    G-protein-coupled receptors. When a ligand activates the receptor,
    a conformation change occurs in the receptor that exchanges GDP
    for GTP on the \ensuremath{\alpha} subunit and this triggers the
    dissociation of the \ensuremath{\alpha} subunit from
    \ensuremath{\beta\gamma} dimer and the receptor. After the free
    \ensuremath{\alpha} unit works on target proteins, GTP will be
    hydrolyzed to GDP. The GTPase activity is enhanced by binding of
    the RGS}.}
\label{fig:G_protein}
\end{figure}

G-protein-coupled receptor's primary function is to transduce
extracellular stimuli into intracellular
signals \cite{Bockaert03,Kobilka06}(Fig.\ \ref{fig:G_protein}). These
receptors' intracellular domains interact with heterotrimeric
G-proteins \cite{Bockaert03}. G-protein-coupled receptors are commanded
both by a wealth of stimuli to which they respond, as well as by the
assortment of intracellular signaling pathways they activate. These
include light, neurotransmitters, odorants, biogenic amines, lipids,
ligands, hormones, nucleotides, and chemokines \cite{Kobilka06}.

G-proteins are composed of \ensuremath{\alpha, \beta,} and
\ensuremath{\gamma} subunits. \ensuremath{\beta and \gamma} form
 the  \ensuremath{\beta\gamma} subcomplex, and the \ensuremath{\alpha}
  subunit can be bound to GDP or GTP. When a ligand activates the
G-protein-coupled receptor, it induces a conformational change in the
receptor that exchanges GDP for GTP on the \ensuremath{\alpha} subunit
and this triggers the dissociation of the \ensuremath{\alpha} subunit,
which is bound to GTP, from the \ensuremath{\beta\gamma} dimer and the
receptor \cite{Bockaert03,Kobilka06}. Then the \ensuremath{\alpha}
subunit will act on the target proteins. After that, the bound GTP
will be hydrolyzed to GDP, and the \ensuremath{\alpha} unit, which is
bound to GDP, will bind the \ensuremath{\beta\gamma} dimer and the
receptor again. From step 5 to step 1(Fig.\ \ref{fig:G_protein}),
hydrolysis from GTP to GDP happens, and is enhanced by binding a
regulator of G-protein signaling(RGS) \cite{Kobilka06,Oldham96}.

While the mechanism of G-protein signaling is highly conserved, the
diversity of signaling targets is immense \cite{Robishaw04}. Thus, the
ability to swiftly model the dynamics of G-protein signaling is
advantageous to both drug development and basic research. 
\subsection{Modeling}
Two recent studies use ODEs to model the receptor and ligand binding
dynamics and the G-protein cycle. Yi. et.\ al.\ \cite{Yi2003} used an
ODEs model together with fluorescent resonance energy transfer (FRET)
measurement to characterize the heterotrimeric G-protein cycle in
yeast. Hao, et.\ al.\ \cite{Hao03} also used an ODEs model to indicate
the role of Sst2, an RGS protein, in G-protein cycling.

In our study, we build four new models using the stochastic
Pi-calculus, stochastic Petri Nets, the Kappa language, and
Bio-PEPA. Furthermore, we show that we obtain consistent results when
compared with the ODEs model of Yi, et.\ al.\ \cite{Yi2003} (Fig.\
\ref{fig:result}). The simulators for stochastic Petri Nets
\cite{mor05} and the Kappa language \cite{DV09}, Cell Illustrator and
Cellucidate, respectively, have an interface which is intuitive for
biologists, and the syntax of Bio-PEPA is close to the chemical
reactions notation and relatively easy to use, unlike SPiM, which
requires understanding of communications primitives and concurrent
processes. In order to hide such communication details we develop a
high level notation that uses terminology directly obtained from
biological processes, and we show how to systematically translate a
process in high level notation into a SPiM process.  (See Section
\ref{section:hln}.). Kahramanogullari. et.\ al.\ \cite{Ozan09} and
Guerriero. et.\ al.\ \cite{Maria07} have provided the narrative
languages for SPiM and Bio-PEPA, respectively. Our high level notation
is an alternative to their narrative languages.
\subsection{Outline}
In Section \ref{sec:Meth}, we show how to implement the model for the
G-protein cycle using five different methods. In Section
\ref{sec:rel}, we compare the simulation results from the different
implementations. In Section \ref{section:hln}, we introduce a high
level notations and the translations from high level notation to
chemical reactions and SPiM code. Section \ref{sec:con} contains our
conclusions and a comparison of the simulation and modeling approaches
of the five methods considered in Section \ref{sec:Meth}.

\section{Method}\label{sec:Meth}
\subsection{ODEs modeling} \label{sec:ODE}
The ODEs modeling approach is a traditional method applied by
biologists to explore dynamic biological processes by numerical
integration. Yi.\ et.\ al.\  \cite{Yi2003} give a computational model
of the G-Protein cycle by means of ordinary differential equations
(ODEs). The individual reactions comprising the key dynamics of
heterotrimeric G-proteins in yeast are represented, along with their
rate constants, in Table \ref{tab:table}.

\begin{table}[!htb]
\centering
\caption{\footnotesize{The reactions used to model the G-protein cycle
    and the corresponding rate constants (rate parameters) for each
    reaction  (G-protein tutorial from www.MathWorks.com).
}}
\begin{center}
\begin{tabular}{ |l|l|l|l|}
\hline
 No. & Reaction & Rate Parameters\\ \hline
1& L + R $\leftrightarrow$ RL&$K_{RL}=2\cdot 10^6M^{-1} \cdot s^{-1}$, $K_{RLm}=1
\cdot 10^{-2}s^{-1}$\\ \hline
2& Gd + Gbg $\rightarrow$ G&$K_{G1}=1s^{-1} $\\ \hline
3 &RL + G $\rightarrow$ Ga + Gbg + RL&$K_{Ga}=1\cdot 10^{-5}s^{-1}$\\ \hline
4& R $\leftrightarrow$ null&$K_{Rd0}=4\cdot 10^{-4}s^{-1}, K_{Rs}=4s^{-1}$\\ \hline
5& RL $\rightarrow$ null&$K_{Rd1}=4\cdot 10^{-3}s^{-1}$\\ \hline
6 & Ga $\rightarrow$ Gd& $K_{Gd}=0.11s^{-1}$\\ \hline
\end{tabular}
\end{center}
\label{tab:table}
\end{table} 

Reaction No.\ 1 is receptor-ligand interaction which corresponds to
the transition from step 2 to step 3 in Fig.\ \ref{fig:G_protein}. R
represents receptor; L represents ligand; and RL represents receptor
and ligand bound together. Reaction No.\ 2 is heterotrimeric G-protein
formation which can be found in the transition from step 1 to step 2
in Fig.\ \ref{fig:G_protein}. Gd represents the deactivated
\ensuremath{\alpha} subunit, which is bound to GDP; Gbg represents the
\ensuremath{\beta\gamma} sub-complex, and G represents
\ensuremath{\alpha} bound to \ensuremath{\beta\gamma} and GDP. Notice
that from step 1 to 2, the \ensuremath{\alpha} subunit also binds to
the receptor. However, the original paper does not consider it as a
species, and consequently the ODEs model does not reflect that binding.
Reaction No.\ 3 is G-protein activation which corresponds to the
transition from step 3 to step 4 in Fig.\ \ref{fig:G_protein}. After
the ligand and receptor interact with each other, the
\ensuremath{\alpha} subunit will be activated and bound to GTP, and it
will dissociate from the \ensuremath{\beta\gamma} sub-complex. Ga
represents the activated \ensuremath{\alpha} subunit, which is bound
to GTP. Reaction No.\ 4 is receptor synthesis and
degradation. Reaction No.\ 5 is receptor-ligand degradation. Reaction
No.\ 6 is G-protein inactivation which corresponds to the transition
from step 5 to step 1 in Fig.\ \ref{fig:G_protein}.

The ODEs of the model are as follows:

\renewcommand{\labelenumi}{\roman{enumi}.}
\begin{enumerate}
\item \emph{d}[R]/\emph{dt}=$-K_{RL}[L][R]+K_{RLm}[RL]-K_{Rd0}[R]+K_{Rs}$;
\item 
\emph{d}[RL]/\emph{dt}=$K_{RL}[L][R]-K_{RLm}[RL]-K_{Rd1}[RL]$;
\item 
\emph{d}[G]/\emph{dt}=$-K_{Ga}[RL][G]+K_{G1}[Gd][Gbg]$;
\item 
\emph{d}[Ga]/\emph{dt}=$K_{Ga}[RL][G]-K_{Gd}[Ga]$;
\item
\emph{d}[Gd]/\emph{dt}=$-K_{G1}[Gd][Gbg]+K_{Gd}[Ga]$;
\item
\emph{d}[Gbg]/\emph{dt}=$K_{Ga}[RL][G]-K_{G1}[Gd][Gbg]$;
\item
\emph{d}[L]/\emph{dt}=$-K_{RL}[L][R]+K_{RLm}[RL]$;
\end{enumerate}

According to Table \ref{tab:table}, derived from the reactions diagram
from Yi.\ et.\ al.\ \cite{Yi2003}, there are seven species in the
model. However, they only describe the first four equations. In order
to have one differential equation for each species, we build the
last three equations. They also provide two conservation
relationships: $[Gbg]=Gt-[G]$ and $[Gd]=Gt-[G]-[Ga]$ which are
satisfied by our simulations. (See Section \ref{sec:rel}).

\subsection{Stochastic Pi-calculus modeling}
Process algebras are formal languages originally designed to model
complex reactive computer systems where heterogeneous agents interact
concurrently exchanging information through communication
channels. Because of the similarities between reactive computer
systems and biological systems, process algebras have recently been
used to model biological systems \cite{CardelliActin09,Priami01}.
% In a process algebra, there are typically two halves of a
% communication, sending and receiving, and communication takes place
% when a sender and a receiver react by exchanging a message. We use the standard notation !ch
% to send a message on channel ch and ?ch to receive a message on
% channel ch. \footnote{Note that it is sufficient to use the simplest
%   form of communication where !ch and ?ch exchange an empty message.}
% \abc{explain delay and reference to Artificial biochemistry paper}

The stochastic Pi-calculus \cite{Priami01} is a process algebra where
stochastic rates are imposed on processes, allowing for more accurate
description of biological systems \cite{CardelliActin09}. A process
can be depicted as a collection of interacting automata with two kinds
of reactions: \textit{delay@r} and \textit{interaction@r on ch}. A
delay is a spontaneous change of state performed by a single automaton
at a specified reaction rate \textit {r}.  Interaction, on the other
hand, consists of a send/receive handshake between two automata over a
channel ch, written !ch for send and ?ch for receive.\footnote{Note
  that it is sufficient to use the simplest form of communication
  where !ch and ?ch exchange an empty message.}  The channel ch has an
associated reaction rate \textit{r}. Reaction rates are used to
determine stochastically the reaction to be executed next
\cite{Artificial}.  For our modeling, we use SPiM,
an implementation of the stochastic Pi-calculus that can be used to
run in-silico simulations displaying the change in the populations of
different species over time \cite{Phillips07}.

\begin{figure}[!htb]
\begin{center}
\includegraphics[scale=0.38]{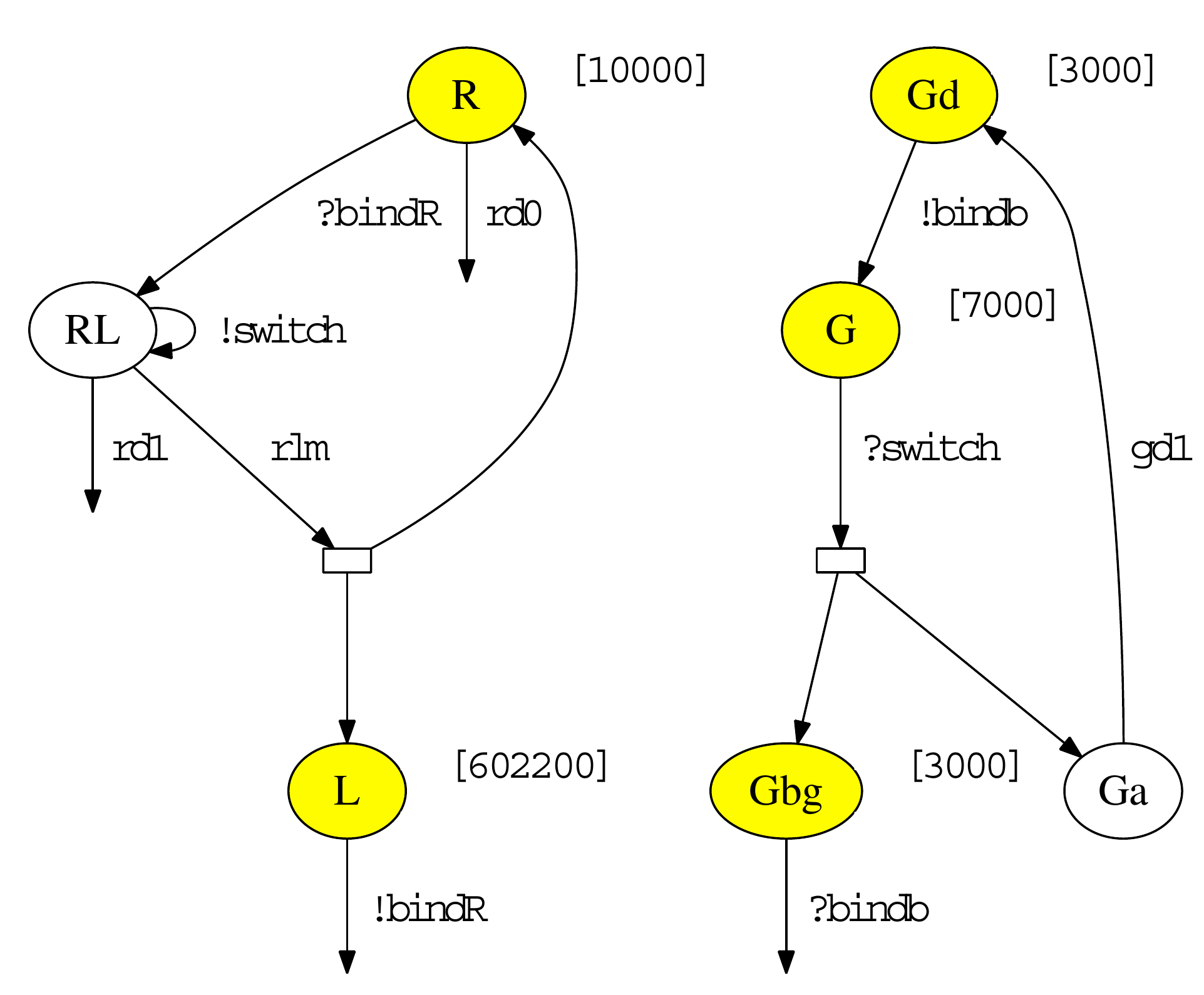}
\end{center}
\caption{\footnotesize{Graphical representation of Pi-calculus modeling}}
\label{fig:SPiM}
\end{figure}

Fig.\ \ref{fig:SPiM} depicts the stochastic Pi-calculus model we build
for the G-protein cycle using SPiM.  We divide the system into two
parts that correspond to two viewpoints. The left hand side is the
viewpoint of ligand and receptor, and the right hand side is the
viewpoint of the G-protein. The initial numbers of different species
are assigned in square brackets. For example, the initial
concentration of R is 10000 and of G is 7000. Notice that RL and Ga
are not initialized. All the reaction rates are also extracted from
the literature \cite{Yi2003} and continuous reactions are converted to
discrete reactions according to standard rules\cite{Cardelli07}. For
example, the pair !bindR/?bindR represents step 2 to 3 in Fig.\
\ref{fig:G_protein}, and it corresponds to reaction No.\ 1 in Table
\ref{tab:table}, and the pair !switch/?switch represents the step 3 to
4 in Fig.\ \ref{fig:G_protein}, and it corresponds to reaction No.\ 3
of Table \ref{tab:table}. The corresponding SPiM code is in Table
\ref{tab:code}. The ``directive plot'' statement declares the species
that will be plotted, in this case RL, G, and Ga. At the end of the
program, each ``run'' statement declares the initial amount for
different species. In the ODEs model, the amount of R is initialized
to be 6.022E17 and the rate of the reaction No.\ 1 in Table
\ref{tab:table} is 3.32E-18 with the assumed reaction volume 1 L
($M=mol \cdot L^{-1}$, $mol\cong 6.022 \times 10^{23}$). Because
  simulating stochastic models is slower than solving ODEs, we
  scale the reaction volume down to 1.0E-12 L in order to speed up the
  stochastic simulation. Given this new reaction volume, the initial
  amount of R is 6.022E5, and
  the new reaction rate is 3.32E-6\cite{Cardelli07}. The other reaction rates remain unchanged, because they are discrete and do not depend on the volume. Assuming that
pressure and temperature are unchanged, 
proportionally scaling  reaction rates and reagent concentrations
preserves the correctness of the modeling.  We do
not model the receptor generation aspect of reaction No.\ 4 in Table
\ref{tab:table} (null $\rightarrow$ R), because on the one hand, this
  reaction is inconsequential to the whole system, as suggested by the
  result of Fig.\ \ref{fig:result}, and on the other hand, the
  stochastic pi-calculus
  can not describe the generation of a process from the empty
  process. Furthermore, the initial amounts of receptor and G-protein guarantee
abundance of receptor to enable the activation of G-protein. More
details about our model can be found in the companion technical report
\cite{G-Protein-SPiM}.
\begin{table}[!htb]
\caption{SPiM code for G-protein Cycle}
  \centering
{\small
\begin{tabular} {lll}
\\
directive sample 600.0& new bindR@rl:chan&and L()=( !bindR(); () )\\
directive plot RL(); G(); Ga()& new switch@ga:chan&\empty\\
\empty &\empty&let Gd()= ( !bindb(); G() )\\
val rl = 3.32E-6&let R()= &and G()= ( ?switch(); (Ga() $\mid$ Gbg() ) )\\ 
val rlm= 0.01& (&and Ga()= ( delay@gd1; Gd())\\
val rs= 4.0&\quad do ?bindR(); RL()&and Gbg()= ( ?bindb(); () )\\ 
val rd0= 4.0E-4&\quad or delay@rd0; ()&\empty\\
val rd1= 4.0E-3& ) &run 602200 of L()\\
val ga= 1.0E-5&and RL()=&run 10000 of R() \\
val gd1= 0.11&(&run 7000 of G()\\
val g1= 1.0&\quad do delay@rlm; ( R() $\mid$ L() )&run 3000 of Gd()\\
\empty& \quad or delay@rd1; () &run 3000 of Gbg()\\
new bindb@ &\quad or !switch(); RL() )&
\\
\hline
\end{tabular}
}
\label{tab:code}
\end{table}

\subsection{Petri Nets modeling}

\begin{figure}
\begin{center}
\includegraphics[scale=0.60]{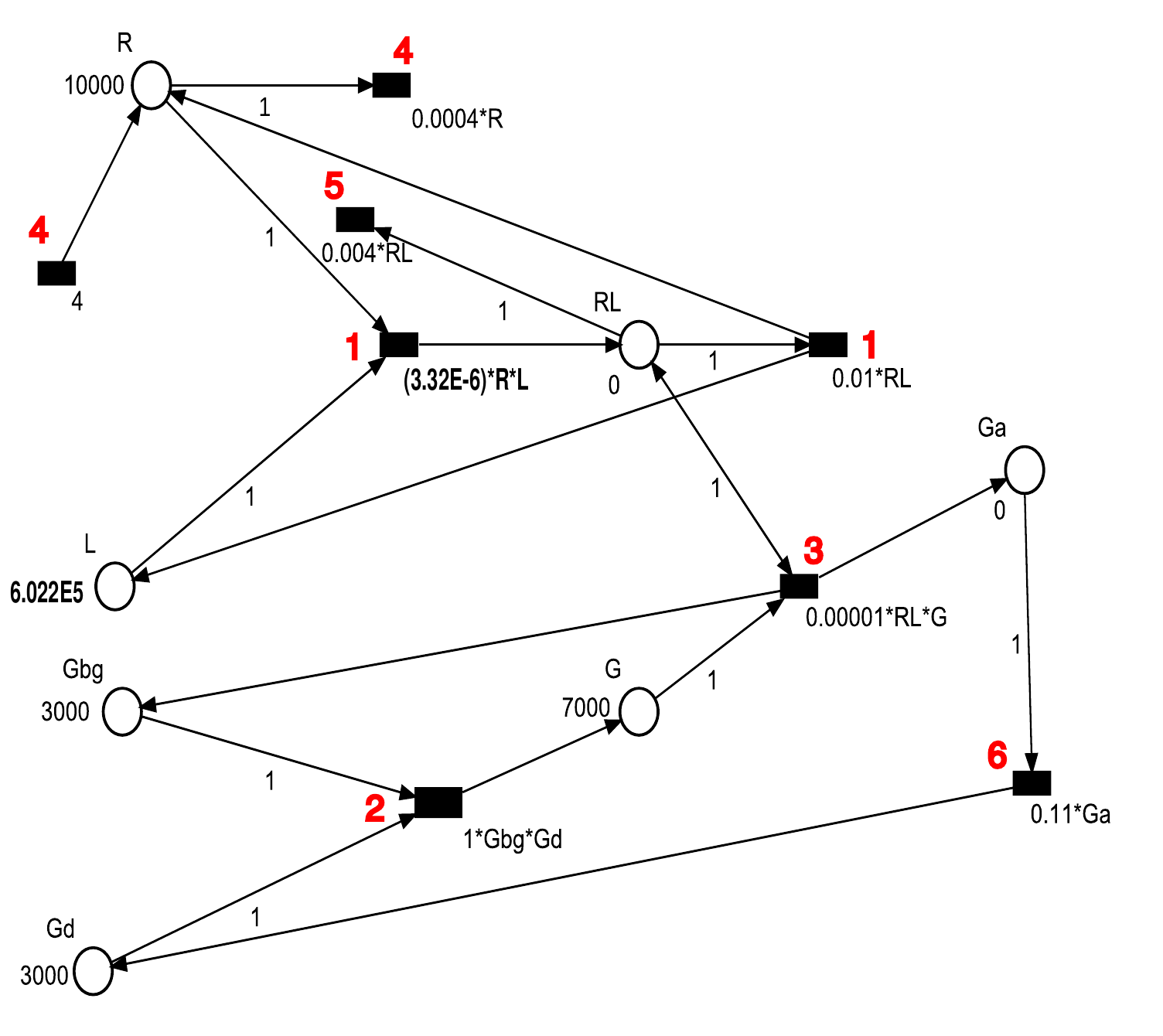}
\end{center}
\caption{\footnotesize{Petri Nets modeling}}
\label{fig:Petri}
\end{figure}

A Petri Net is a mathematical concept developed in the early 1960s by
Carl Adam Petri to describe discrete distributed systems. Petri Nets are suitable for
modeling and analyzing the dynamics of concurrent systems whose
behavior could be described by finite sets of atomic processes and
atomic states  \cite{mor05}. Petri Nets have subsequently been adapted
and extended in many directions, including quantitative analysis of
biological networks  \cite{JW03}.

The basic Petri Net is a directed bipartite graph with two kinds of
nodes which are either places or transitions and directed arcs which
connect nodes. In modeling biological processes, place nodes represent
molecular species and transition nodes represent reactions. We use
Cell Illustrator to develop our model of the G-protein cycle (\href{http://www.cellillustrator.com}
{\texttt{www.cellillustrator.com}}).

Fig.\ \ref{fig:Petri} is the Petri Net we build to model the G-protein
cycle. Place nodes are represented as circles, and transition nodes
are represented as boxes. The arcs are labeled with an integer weight,
which represents the minimum value of input entities needed for the
reactions to happen. The initial number of species and reaction rates
can be read from the graph. For example, the initial amount of
Receptor is 10000 and the initial amount of \ensuremath{\beta\gamma}
sub-complex is 3000, while Ga and RL are initialized to 0. Here we
scale the reaction volume to be 1.0E-12 L, which is the same as in the
SPiM model. The numbers in red correspond to the No.\ column in
Table \ref{tab:table}.

\subsection{Kappa language modeling}

\begin{table}[!htb]
\caption{\footnotesize{G-protein cycle is represented in Kappa
    language.}}
\centering
{\small
\begin{center}
\begin{tabular}{|l|l|}
\hline
No. & Kappa Statement\\ \hline
1&R(r), L(l) $\leftrightarrow $ R(r!1), L(l!1) @ 3.32e-06,0.01\\ \hline
2&Gbg(bg), Gd(alpha $\sim$ GDP,a) $\rightarrow$ Gbg(bg!1), Gd(alpha
$\sim$ GDP,a!1) @ 1.0\\ \hline
3&Gbg(bg!1), Gd(alpha $\sim$ GDP,a!1), R(r!2), L(l!2) $\rightarrow$
Gbg(bg),\\\empty &Gd(alpha $\sim$ GTP,a), R(r!1), L(l!1) @ 1.0e-05\\\hline
4&R(r)$\leftrightarrow$ @4e-04,4\\ \hline
5&L(l!1), R(r!1) $\rightarrow$ @ 0.004\\ \hline
6&Gd(alpha $\sim$ GTP,a) $\rightarrow$ Gd(alpha $\sim$ GDP,a) @ 0.11\\ \hline
\end{tabular}
\end{center}
}
\label{tab:kappa}
\end{table}

Kappa is a formal language for defining agents (typically meant to
represent proteins) as sets of sites that constitute abstract
resources for interaction. It is used to express rules of interactions
between proteins characterized by discrete modification and binding
states \cite{DV09}. The Kappa language modeling platform Cellucidate
is available at \href{http://www.cellucidate.com}
{\texttt{www.cellucidate.com}}.

In the Kappa language,  reaction rules are described by rewriting rules
between lists of agents. Each agent has a name and binding sites. For
example, in R(r), R is the agent and r is its binding site. Agents can
become bound, and the two end points of a link between two agents is
indicated by !i, for some index i.
$\sim$value specifies the internal state of a site on the
agent, $\leftrightarrow$ specifies a bidirectional
reaction, $\rightarrow$ specifies an unidirectional reaction, and @value
specifies the reaction rate.  For example, {\small
\begin{center}
R(r), L(l) $\leftrightarrow $ R(r!1), L(l!1) @ 3.32e-06,0.01  
\end{center}
}
is a bidirectional reaction, where R and L are agent names, r is the
binding site of R, l is the binding site of L; 0.01 is the reaction
rate from left to right, and 3.32e-06 is the reaction rate from right
to left. In R(r!1) and L(l!1), 1 is the index of the link that binds R
and L at their binding sites. In our G-protein example, it corresponds
to the binding of receptor and ligand. Gd(
  \ensuremath{\alpha}$\sim$GDP,a) represents the deactivated
  G-protein. $\sim$GDP represents the internal state of the
  \ensuremath{\alpha} subunit where it is bound to GDP.

\begin{figure}
\begin{center}
\includegraphics[scale=0.6]{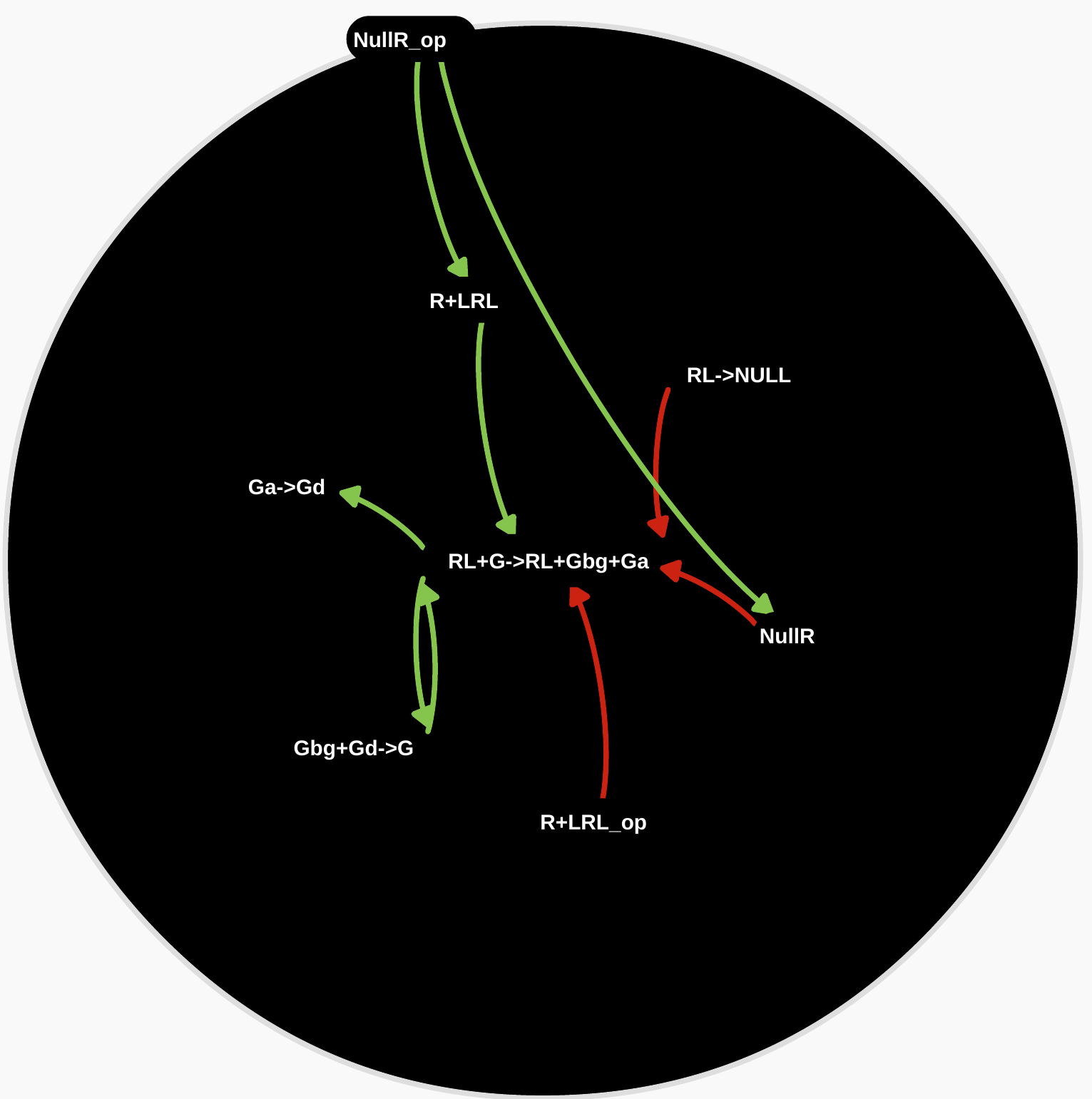}
\end{center}
\caption{\footnotesize{Graphical representation of Kappa language modeling}}
\label{fig:Kappa_fig}
\end{figure}

The Kappa language processes for the G-protein cycle are described in
Table \ref{tab:kappa}. The statements correspond to the reactions and
rates in Table \ref{tab:table}\footnote{Note that the initial amount
  of different species can be set in Cellucidate.}.  The initial
number of species can not be read from Table \ref{tab:kappa}, but in
fact, we use the same number as the Petri Nets model and SPiM model. For
example the initial number for ligand is 6.022E5. Fig.\
\ref{fig:Kappa_fig} is the graphical representation of the processes
described in Table \ref{tab:kappa} automatically generated by
Cellucidate. $R+LRL$ means $R+L\rightarrow RL$, and $R+LRL\_op$ is the
opposite direction: $R+L\leftarrow RL$. $NullR$ means $R\rightarrow
null$, and $NullR\_op$ is the opposite direction: $R\leftarrow
null$.  The reactions with red arrows will consume \emph{R} or
\emph{RL}, having a negative effect on the reaction $R+RL\rightarrow Ga +
Gbg + RL$, while the green arrows have positive effects.

\subsection{Bio-PEPA modeling}

Bio-PEPA\cite{Biopepa09} is an extension of PEPA\cite{Pepa96}, which
is specifically designed for modeling biochemical networks by
explicitly defining the stoichiometry of reactions. The two existing
tools for working with Bio-PEPA models are the Bio-PEPA Workbench and
the Bio-PEPA Eclipse Plug-in\cite{tool09}. We use the Bio-PEPA Eclipse
Plug-in to model the G-protein Cycle, shown in Table
\ref{tab:pepa}. In Table \ref{tab:pepa}, Parameter Definitions consist
of rate constant declarations; Functional Rates are declarations of
reaction names and the corresponding reaction rates. For example, in
$kineticLawOf bindRL : fMA(rl)$, the name of the reaction is $bindRL$,
the function for the law of Mass-Action is denoted by $fMA$, and $rl$
is the rate constant; under Species Components there is a
representation of the reactions matching the corresponding
differential equations of the ODEs model. For example,
\[
G = bindAB >> G + activ << G
\] 
indicates the role of reagent G in the related reactions. $bindAB>>G$
means $G$ is produced in reaction $bindAB$ and $activ<<G$ means $G$ is
consumed in reaction $activ$. It also corresponds to \\
\begin{center}
\emph{d}[G]/\emph{dt}=$-K_{Ga}[RL][G]+K_{G1}[Gd][Gbg]$,\\
\end{center}
where $g1$ is the rate constant of reaction $bindAB$, and
$ga$ is the rate constant of reaction $activ$, corresponding to
$K_{G1}$ and $K_{Ga}$ respectively.

  Finally, the initial concentrations for each species are declared
  under Model Components.  Because in our example the reactions are in
  the same cell, and the locations are the same for all species, the
  default location is used and left implicit.

   The reaction rates and initial concentrations are all
  consistent with the other three models we built.
\begin{table}[!htb]
\caption{Bio-PEPA code for G-protein Cycle}
  \centering
{\small
\begin{tabular} {l}
\\
// Parameter Definitions\\
 $rl = 3.32E-6;$\\
 $rlm= 0.01;$\\
 $ga= 1.0E-5;$\\
 $gd1= 0.11;$\\
 $g1= 1.0;$\\
 $rd1= 4.0E-3;$\\
 $rs= 4.0;$\\
 $rd0= 4.0E-4;$\\\\
// Functional Rates\\
$kineticLawOf bindRL : fMA(rl);$\\
$kineticLawOf bindAB : fMA(g1);$\\
$kineticLawOf disRL 	: fMA(rlm);$\\
$kineticLawOf activ 	: fMA(ga);$\\
$kineticLawOf hydro 	: fMA(gd1);$\\
$kineticLawOf degRL  : fMA(rd1);$\\
$kineticLawOf degR   : fMA(rd0);$\\
$kineticLawOf proR   :  rs;$\\\\
// Species Components\\
$L = disRL >> L + bindRL << L;$\\
$R = disRL >> R + bindRL << R + degR << R + proR >> R;$\\
$RL = disRL << RL + bindRL >> RL + activ (+) RL + degRL <<RL;$\\
$G = bindAB >> G + activ << G;$\\
$Ga = activ >> Ga + hydro << Ga;$\\
$Gd = bindAB << Gd + hydro >> Gd;$\\
$Gbg = bindAB << Gbg + activ >> Gbg;$\\\\
// Model Components\\
$L[602200]<*>R[10000]<*>RL[0]<*>G[7000]$\\
$<*>Gd[3000]<*>Gbg[3000]<*>Ga[0]$\\
\hline
\end{tabular}
}
\label{tab:pepa}
\end{table}

\section{Simulation Results and Comparison}\label{sec:rel}
\begin{center}
\begin{figure}[htbp]
\begin{tabular}{cc}
   \includegraphics[scale=0.40]{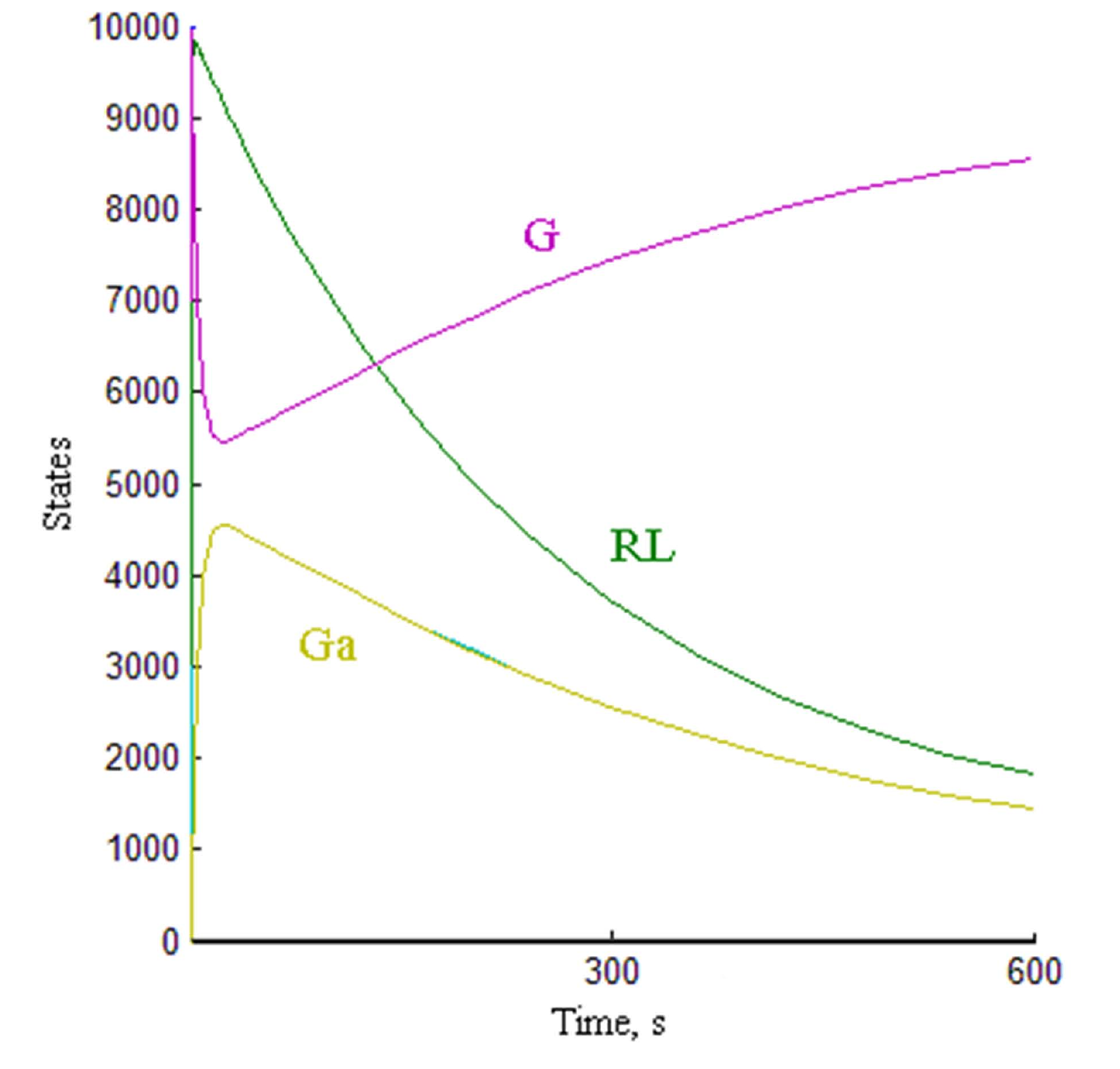}&
    \includegraphics[scale=0.12]{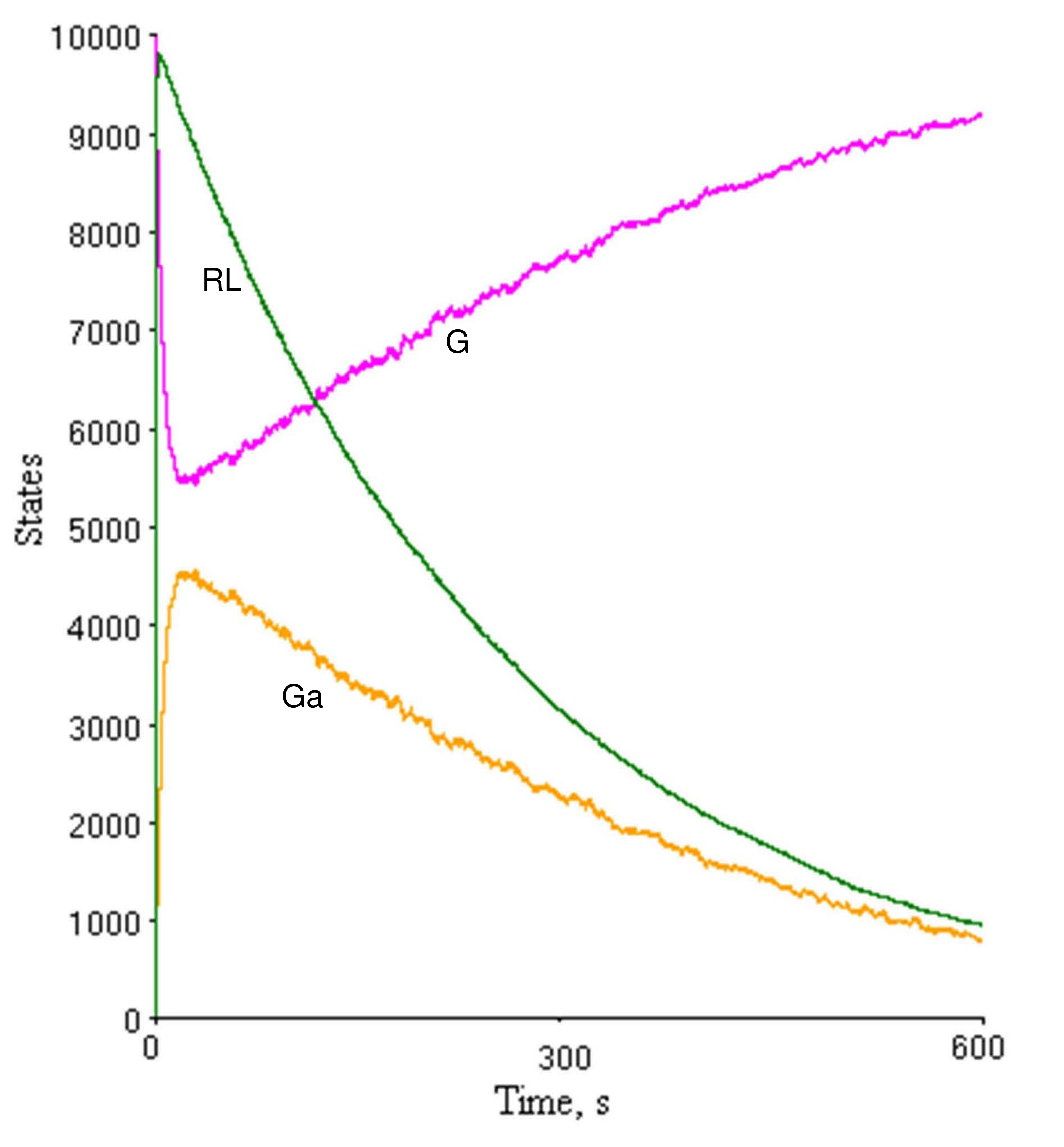}\\
   (a)&(b)\\
    \includegraphics[scale=0.30]{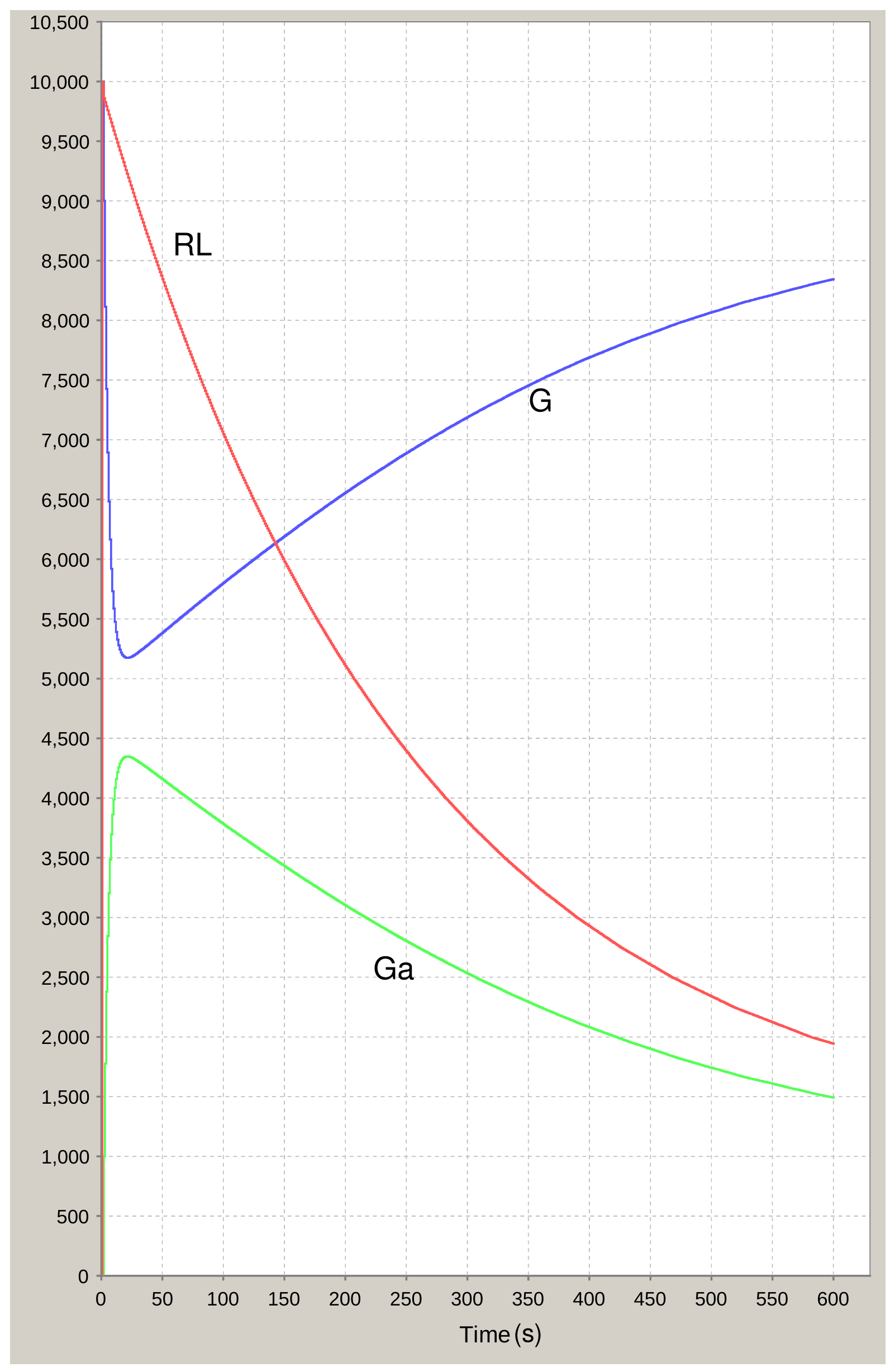}&

    \includegraphics[scale=0.25]{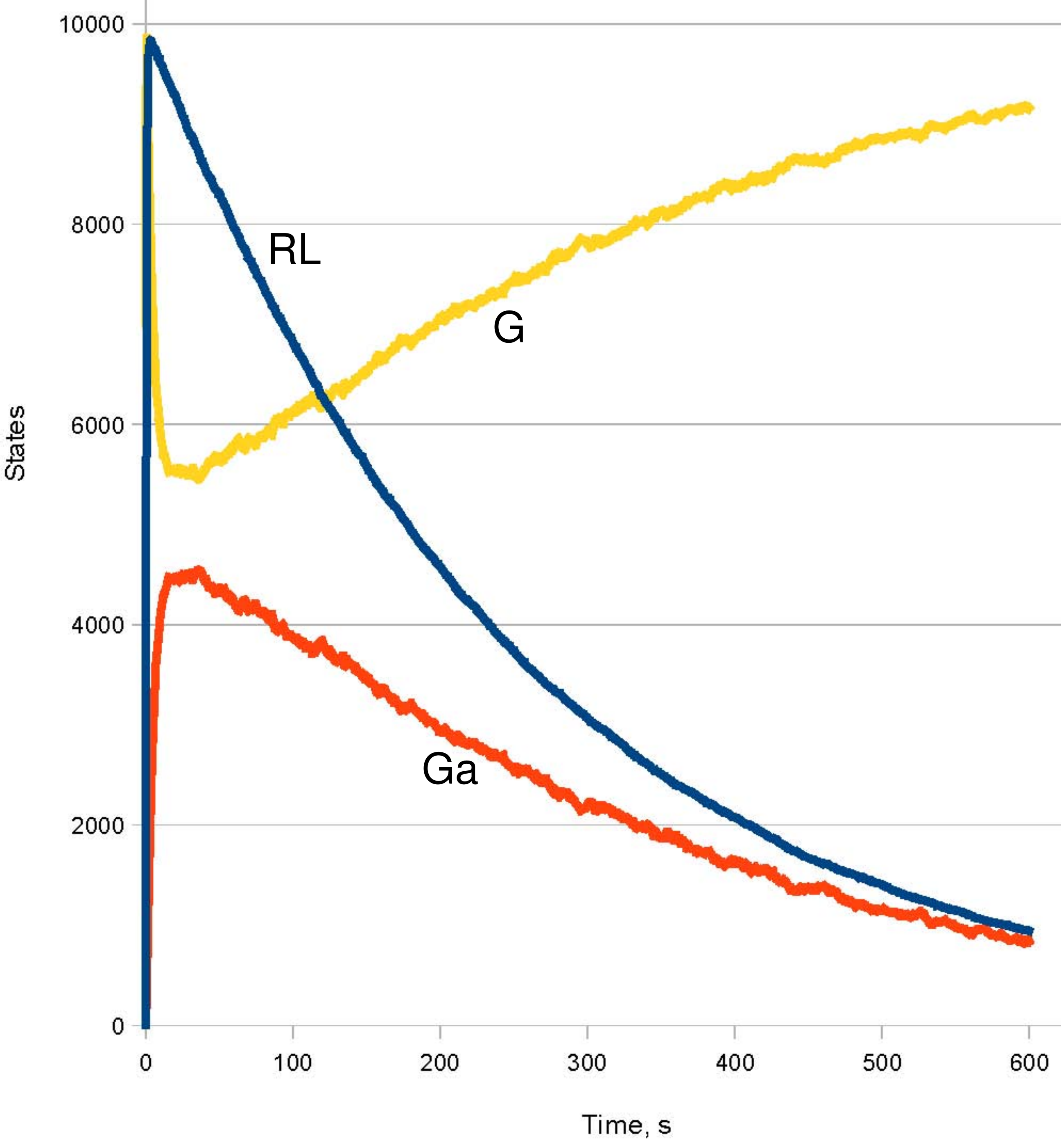}\\ 
    (c)&(d)\\
    \multicolumn{2}{c}{\includegraphics[scale=0.45]{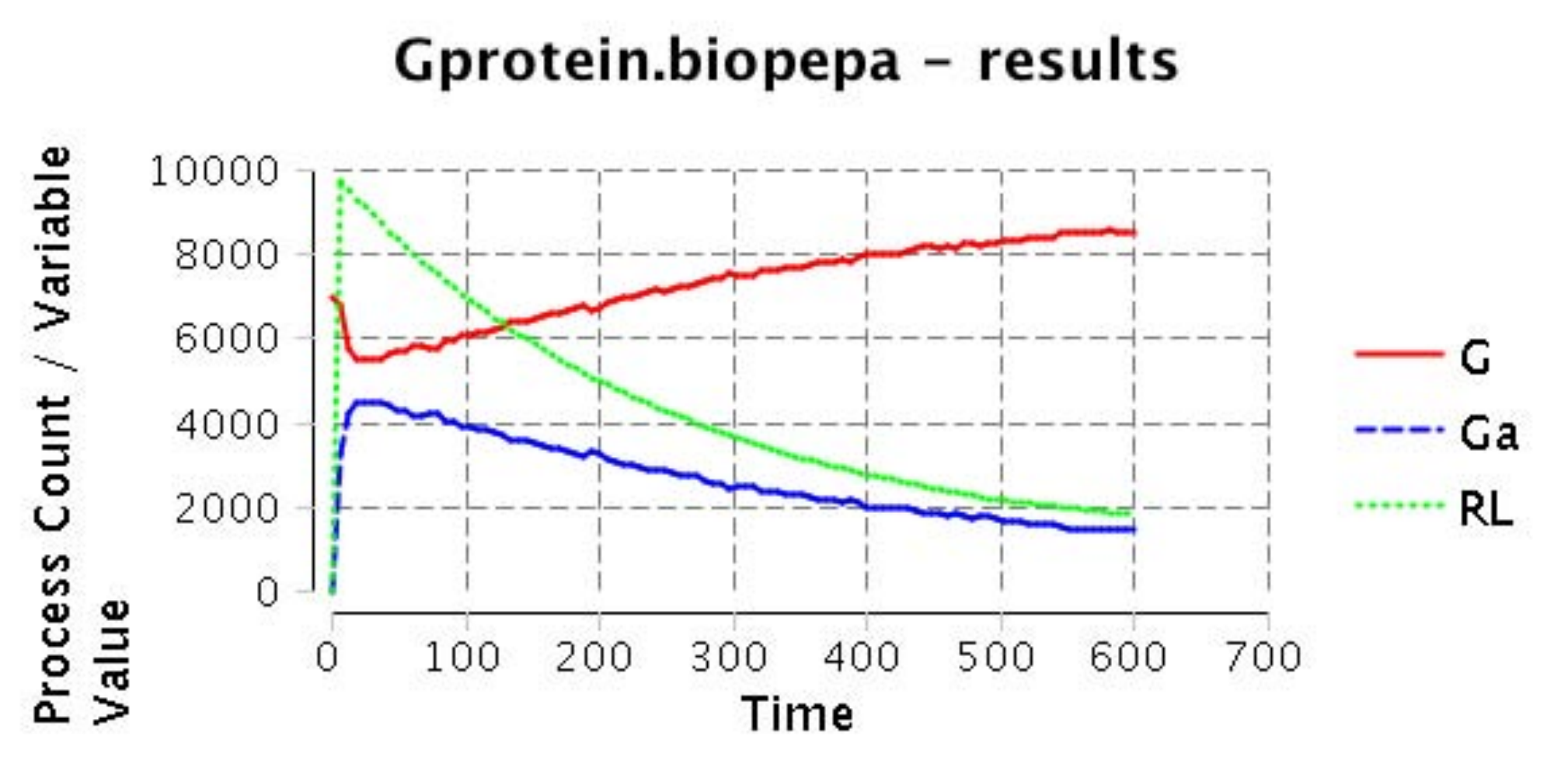}}\\
    \multicolumn{2}{c}{(e)}
\end{tabular}
  
\caption{\footnotesize{(a). Simulation result of ODEs model from MathWorks.
  (b). Simulation result of process model from SPiM. (c). Simulation result of Petri Nets model from Cell
    Illustrator. (d). Simulation result of Kappa language from
    Cellucidate. (e). Simulation result of Bio-PEPA from
    Eclipse Plug-in. In all five cases, RL represents receptor bound to
    ligand, Ga represents the activated G-protein, and G represents
    the deactivated G-protein. The x axis is time in seconds, and the y axis is intensity of
    states. }}
\label{fig:result}
\end{figure}
\end{center}

Although we are not proposing a theoretical comparison of the five
modeling techniques, or the five implementations in different
formalisms, we show here consistent experimental simulation
results. SPiM, Cell Illustrator, Cellucidate, and Bio-PEPA all
  use Gillespie's algorithm for simulation. Fig.\ \ref{fig:result}
shows that the results from those four simulations are consistent with
the result of the original ODEs modeling. The plots show that RL is
consumed as the reaction proceeds. The curves of G and Ga decline and
increase oppositely, and the sum of these two species is a constant,
which is equal to the initial amount of G-protein, so the conservation
relationships mentioned by Yi.\ et.\ al.\ are confirmed
($[Gbg]=Gt-[G]$, $[Gd]=Gt-[G]-[Ga]$) \cite{Yi2003}.

These computational models can be classified according to different
properties.

\textbf{Simulation Approach.}  The simulation approach can be
stochastic or deterministic. While the deterministic approach is
faster, the stochastic approach is more accurate. In particular,
because of the randomness of dynamic biological systems, the
stochastic approach can make the simulation of biological processes
more faithful. Furthermore, while a deterministic system may get stuck
into a specific state, stochastic noise allows the system to oscillate
in and out of that state \cite{vilar02}.

Our four models of the G-protein cycle in the stochastic
Pi-calculus, stochastic Petri Nets, the Kappa language, and Bio-PEPA fall into
the category of stochastic simulation, while the traditional ODEs
model is deterministic.

Far from competing with each other, both modeling approaches are
complementary, and often it is possible to derive one model from the
other. In particular, Feret, et.\ al.\ show how to convert a
rule-based model, such as a Kappa model, into a reduced systems of
differential equations \cite{DV09}, Priami, et.\ al.\ show how to translate
an ODEs model into a Stochastic Pi-calculus model \cite{Priami09},
and Cardelli, et.\ al.\ show how to convert processes to ODEs \cite{Cardelli08}.

\textbf{Modeling Approach.}  The five models discussed here are either
chemical reaction centric or process centric. The ODEs, Petri Nets,
Kappa, and Bio-PEPA models are chemical reaction centric. In
particular, Kappa takes a rule based approach that corresponds to high
level chemical reactions. For each chemical reaction there is an
equation, a transition or a rule in the corresponding model. On
the other hand, the stochastic Pi-calculus model is process centric;
there is one process for each component or observable product, leading
to a much simpler model. Andrew Phillips shows the simplicity of the
SPiM process model in contrast with the chemical reactions model
\cite{Phillips09}.

While ODEs are low level and close to chemical reactions, Petri Nets
constitute a graphical representation of chemical reactions. A SPiM
model is a low level representation of individual protein behavior
using send and receive communication primitives over a channel, and
Kappa is a graphical high level abstraction of chemical reactions.

\vspace{-30.0pt}
\section{High Level Notation} \label{section:hln}

\newcommand{\action}[1]{$\mbox{\textbf{#1}}$} 
\newcommand{\defB}[7] { \mbox{SPiM} (\action{#1}) = \{#2 \rightarrow^{#3} #4 ; #5\rightarrow^{#6}#7\} }
\newcommand{\defD}[4] { \mbox{SPiM} (\action{#1}) = \{#2 \rightarrow^{#3} #4\}}
\newcommand{\defDe}[4] { \mbox{SPiM} (\action{#1}=  \{#2 \rightarrow^{#3} #4\}}
\newcommand{\defA}[8]{\mbox{SPiM} (\action{#1}) = \{#2
  \rightarrow^{#3} #4\ \mid \ #8\ ; #5\rightarrow^{#6}#7\}}

\newcommand{\bind}[5] {\{#1 \rightarrow^{!#5@#4} #3 ; #2\rightarrow^{?#5@#4}()\}}
\newcommand{\dis}[4]{\{#1 \rightarrow^{delay@{#4}}#2\ \mid \ #3\}}
\newcommand{\degrad}[3]{\{#1 \rightarrow^{delay@{#3}}#2\}}
\newcommand{\acd}[6]{\{#1 \rightarrow^{?#6@#5} #3\ \mid \ #4\ ; #2\rightarrow^{!#6@#5}#2\}}
\newcommand{\trans}{\rightarrow_{eval}}
\newcommand{\state}[2]{#1 \cup \mbox{\textbf{{SPiM}(#2}}}

Both ODEs and Petri Nets correspond closely to chemical reactions, and
for the average biologist, they are relatively easy to understand.
However, according to our experience in the classroom and in the lab,
communicating with biology students and scientists, understanding the
stochastic Pi-calculus, the Kappa language and Bio-PEPA (to a lesser
extent) to be able to model molecular processes, requires a
considerable initial effort.  While Kappa has a user friendly
graphical syntax available through Cellucidate. SPiM, on
the other hand, still needs encoding in the stochastic Pi-calculus. To address this issue and enable modeling
using descriptions directly obtained from biological processes,
a narrative language has been proposed for SPiM
\cite{Ozan09}. Here we propose an alternative high level
notation, and we show how to systematically translate it into
SPiM programs and also chemical reactions. 
%Furthermore, we show how to
%translate arbitrary higher order reactions.

\begin{table}[!htb]
\caption{High Level Notation}
\centering

\begin{tabular}{c}
{\small
\begin{tabular} {llllll}
\\
Action A &::= &  \action{bind(a, b, c, r)} &
&$\mid$& \action{dimerize(a, b, c, r)}\\
&$\mid$&\action{activate(a, b, c, r)}&
&$\mid$&\action{activateAnddissociate(a, b, c, d, r)}\\
&$\mid$& \action{phosphorylate(a, b, c, r)}&
&$\mid$& \action{dissociate (a, b, c, r)}\\
&$\mid$& \action{degrade(a, r)}&
&$\mid$&\action{hydrolyze(a, b, r)}\\
%&$\mid$&\action{reaction(a1,a2...,am,b1,b2...,bn,m,n,r )}&
%&  &  \\
Process P &::=& A; P $\mid$ ()\\ \\
\end{tabular}
}
\\
{\small
\begin{tabular}{ll}
  $\defB{bind(a, b, c, r)} {a} {!ch@r} {c} {b} {?ch@r}
  {()}=\mbox{SPiM} (\action{dimerize(a, b, c, r)})$&
where ch is new\\
$\defB{phosphorylate(a, b, c, r)} {a} {!ch@r} {c} {b} {?ch@r} {b}$&
where  ch is new \\
$\defB{dimerize(a, b, c, r)} {a} {!ch@r} {c} {b} {?ch@r} {()}=\mbox{SPiM} (\action{bind(a, b, c, r)})$&
where  ch is new \\
$\defB{activate(a, b, c, r)}{a}{!ch@r}{c}{b}{?ch@r}{b}$&
where  ch is new \\
$\defA{activateAnddissociate(a, b, c, d, r)}{a}{!ch@r}{c}{b}{?ch@r}{b}{d}$&
where  ch is new \\
$\defD{dissociate(a, b, c, r)}{a}{delay@r}{b  \hskip 5 pt \mid  \hskip 5 pt c}$\\
$\defDe {hydrolyze(a, b, r)}{a}{delay@r}{b}$\\
$\defDe{degrade(a, r)}{a}{delay@r}{()}$\\\\
%$\mbox{SPiM}(\action{reaction(a1,a2...am,b1,b2...,bn,m,n,r
 % )})=\{a1\rightarrow^{!ch1@r}();$&
%where ch1, ch2...ch(m-1)\\
%$a2\rightarrow^{?ch1@r;!ch2@r}();...am\rightarrow^{?ch(m-1)@r}b1\mid b2 \mid
%... \mid bn\}$& are all new\\\\
\end{tabular}
}
\\
{\small
\begin{tabular}{lll}
  SPiM* (()) &=& $\emptyset$\\
  SPiM* (A; P) &=& SPiM(A) $\cup$ SPiM* (P)\\

\end{tabular}
}
\\
\hline
\end{tabular}
\label{tab:hln}
\end{table}

\begin{table}[!htb]
\caption{Translation of High Level Notation into Chemical Reactions}
\centering
\begin{tabular}{c}
{\small
\begin{tabular}{l}
\\
$ \mbox{Reaction} (\action{bind(a, b, c, r)}) =  a + b \rightarrow^r c$\\
$ \mbox{Reaction} (\action{phosphorylate(a, b, c, r))}) =  a + b
\rightarrow^r c + b$\\
$\mbox{Reaction} (\action{dimerize(a, b, c, r)}) = a + b \rightarrow^r
c$\\
$\mbox{Reaction} (\action{activate(a, b, c, r)}) =  a + b
\rightarrow^r c +b$\\
$\mbox{Reaction} (\action{activateAnddissociate(a, b, c, d, r)}) =  a
+ b \rightarrow^r c + d+ b$\\
$\mbox{Reaction} (\action{dissociate(a, b, c, r)}) =a \rightarrow^r b
+ c$\\
$\mbox{Reaction} (\action{hydrolyze(a, b, r))}) =a \rightarrow^r b$\\
$\mbox{Reaction} (\action{degrade(a, r)}) = a \rightarrow^r null$\\\\
%$\mbox{Reaction} (\action{reaction(a1,a2...,am,b1,b2...,bn,m,n,r
 % )}) = a1+a2+...+am\rightarrow^r b1+b2+...+bn$\\\\
\end{tabular}
}
\\
{\small
\begin{tabular}{lll}
  Reaction* (()) &=& $\emptyset$\\
  Reaction* (A; P) &=& Reaction(A) $\cup$ Reaction* (P)\\
\end{tabular}
}
\\
\hline
\end{tabular}
\label{tab:rec}
\end{table}

We write a, b, c, d, etc.\ for species, r for reaction rate constants, A for
actions, and P for a process consisting of a possibly empty sequence
of actions. Table \ref{tab:hln} shows a list of actions representing
familiar biological processes. 
\textbf{bind(a, b, c, r)} means two reagents \textbf{a} and \textbf{b}
bind together to generate the product \textbf{c} at rate \textbf{r}.
Similarly, \action{dimerize(a, b, c, r)} means that \textbf a and
\textbf b form a dimer denoted by \textbf c at rate \textbf r.
\textbf{dissociate(a, b, c, r)} is the opposite action to
\textbf{bind(a, b, c, r)}; \textbf{a} is broken down into \textbf{b}
and \textbf{c} at rate \textbf{r}.
\textbf{activate(a, b, c, r)} and \textbf{activateAnddissociate(a, b,
  c, d, r)} basically represent the same biological process where the
reagent \textbf{a} becomes activated in the presence of another
reagent \textbf{b} at rate \textbf{r}. The difference is that
\textbf{activate} only generates the activated state product
\textbf{c} while \textbf{activateAnddissociate} generates the
activated product, and the product simultaneously dissociates into two
products \textbf{c} and \textbf{d}.
\textbf{phosphorylate(a, b, c, r)} represents phosphorylation at 
rate \textbf{r}, where \textbf{a} is the protein reagent, \textbf{b} is the
kinase, and \textbf{c} is the product.
\textbf{hydrolyze(a, b, r)} represents the hydrolysis at rate
\textbf{r}, where \textbf{a} is the reagent and \textbf{b} is the
product.
\textbf{degrade(a, r)} means \textbf{a} degrades at rate \textbf{r}.
%%
%\textbf{reaction(a1,a2,...,am,b1,b2,...bn,m,n,r)} is used to
%represent an arbitrary higher order reaction. a1, a2..., am are the reactants
%and b1, b2..., bn are the products. m is the number of reactants, n is
%the number of products and r is the rate constant.
%
Notice that \textbf{bind(a, b, c, r)} and \action{dimerize(a, b, c, r)}
are basically the same operation, but in order to distinguish these two
different biological processes, we have two names. In biology,
dimerization is the binding which can generate a dimer.

Table \ref{tab:hln} also defines the translation function SPiM,
that given an action in high level notation returns the
corresponding SPiM process, and SPiM*, its homomorphic extension to
processes. To aide readability, instead of using stochastic
Pi-calculus code, we use the equivalent graph representation of SPiM
processes \cite{CardelliGTP09}. Similarly, Table \ref{tab:rec} defines
the translation functions Reaction and Reaction* that translate
processes in high level notation into chemical reactions.

%%%%%%%%%%%%%%%%
% Table \ref{tab:hln} shows some actions representing familiar
% biological processes.  \action{\small bind(a, b, c, r)} means that
% \textbf a and \textbf b bind to become \textbf c at rate \textbf
% r. Similarly, \action{\small dimerize(a, b, c, r)} means that  \textbf
% a and \textbf b form a dimer denoted by \textbf c at rate \textbf
% r.
% %
% \abc{explain how to read activate}
% %
% \action{\small activateAnddissociate(a, b, c, d, r)} is motivated by
% the ODE model of Section \ref{sec:ODE} where \textbf a is a
% complexation of \textbf c and \textbf d, and the coexistence with
% \textbf b cause the activation of \textit a. The activation causes
% \textit b and \textbf c to dissociate, and the entire reaction has
% rate \textbf r. In \action{\small phosphorylate(a, b, c, r)} \textbf a
% is a protein and \textbf b is a phosphate. The phosphorylated \textit
% a is called \textbf c, and again \textbf r is the rate. \action{\small
%  hydrolyze} and \action{\small degrade} are self explanatory.
% %%%%%%%%%%%%%%%

Going back to our G-Protein example in Fig.\ \ref{fig:G_protein},
\textbf { bind(Gd, Gbg, G, 1.0)} corresponds to step 1 to 2,
\textbf{ bind(R, L, RL, 3.32e-6)} to step 2 to 3,
\textbf{ activateAnddissociate(G, RL, Ga, Gbg, 1.0e-5)} to the activation
in steps 2 to 3 and the dissociation in step 3 to 4,
\textbf{ dissociate(RL, R, L, 0.01)} represents step 4 to 5 and
\textbf{ hydrolyze(Ga, Gd, 0.11); degrade(R, 4e-4); degrade(RL, 4e-3)}
correspond to step 5 to 1 completing the cycle. 

The translation of the high level notation of the G-protein cycle into
SPiM is as follows:

{\small
\[
 \begin{array}{l}
\textbf {SPiM*(bind(Gd, Gbg, G, 1.0); bind(R, L, RL, 3.32e-6);}\\ 
 \textbf{activateAnddissociate(G, RL, Ga, Gbg,1.0e-5);  dissociate(RL, R, L, 0.01);}\\
 \textbf{hydrolyze(Ga, Gd, 0.11); degrade(R, 4e-4); degrade(RL, 4e-3))}\\
 =
\\

 \bind{Gd}{Gbg}{G}{1.0}{bindb} \cup \textbf {SPiM*(bind(R, L, RL, 3.32e-6);}\\ 
 \textbf{activateAnddissociate(G, RL, Ga, Gbg,1.0e-5);  dissociate(RL, R, L, 0.01);}\\
 \textbf{hydrolyze(Ga, Gd, 0.11); degrade(R, 4e-4); degrade(RL,
   4e-3))}\\
=
\\
\quad$\vdots$
\\
=
  \\
  \bind{Gd}{Gbg}{G}{1.0}{bindb} \cup \bind R L {RL} {3.32e-6} {bindR} \cup\\
  \acd {G} {RL} {Ga}{Gbg} {1.0e-5}{switch}  \cup \dis{RL} R L{ 0.01} \cup \\
  \degrad{Ga}{Gd}{0.11} \cup \degrad{R}{()}{4e-4} \cup \degrad{RL}{()}{4e-3}  
  \end{array}
\]
} 
Finally, the translation of the high level notation of the G-protein cycle
into chemical reactions is as follows: 
{\small
\[
 \begin{array}{l}
\textbf {Reaction*(bind(Gd, Gbg, G, 1.0); bind(R, L, RL, 3.32e-6);}\\ 
 \textbf{activateAnddissociate(G, RL, Ga, Gbg,1.0e-5);  dissociate(RL, R, L, 0.01);}\\
 \textbf{hydrolyze(Ga, Gd, 0.11); degrade(R, 4e-4); degrade(RL, 4e-3))}\\
 =
\\

 Gd + Gbg \rightarrow^{1.0} G \cup \textbf {Reaction*(bind(R, L, RL, 3.32e-6);}\\ 
 \textbf{activateAnddissociate(G, RL, Ga, Gbg,1.0e-5);  dissociate(RL, R, L, 0.01);}\\
 \textbf{hydrolyze(Ga, Gd, 0.11); degrade(R, 4e-4); degrade(RL,
   4e-3))}\\
=
\\
\quad$\vdots$
\\
=
  \\
   Gd + Gbg \rightarrow^{1.0} G \cup R + L \rightarrow^{3.32e-6} RL \cup\\
  G + RL \rightarrow^{1.0e-5} Ga + Gbg + RL  \cup RL
  \rightarrow^{0.01} R + L \cup \\
  Ga \rightarrow^{0.11} Gd \cup R \rightarrow^{4e-4} null \cup  RL \rightarrow^{4e-3} null
  \end{array}
\]
}

Kahramanogullari. et.\ al.\ \cite{Ozan09}, propose a narrative
language that is translated into a model with three kinds of sentences
(association, dissociation and transformation), where species have
explicit binding sites (as in Kappa). They also show how to translate
models into SPiM and how to obtain a model using the narrative
language. Our high level notation is an alternative to their narrative
language, however we translate it directly into SPiM, and we do not need a
notion of model. In their example of Appendix A, the generated code
for processes FcR2 and FcR3, for example, does not correspond to any species in
the biological example being modeled.

\section{Conclusion}\label{sec:con}

Although building computational models of cellular processes is not
new, to the best of our knowledge, this is the first time that a model
has been implemented in five different formalisms.  Furthermore, our
classroom experience shows that this survey paper is a valuable
tutorial introduction to bio-modeling in these diverse
frameworks. In order to make a formal comparison
for all five formalisms, one would need to define translation
functions between the different formalisms and a notion of correctness
for each translation. 
%Although attractive as a future research
%direction, the time between the presentation at the meeting in Jena
%and the deadline for submission was not enough to reach such an
%ambitious goal.

Starting from an exiting ODEs model of the activation cycle of
G-proteins by G-protein coupled receptors, we construct four
simulations in four different formalisms: Stochastic Pi-Calculus,
Stochastic Petri Nets, Kappa, and Bio-PEPA. We also show how to scale
initial concentrations and reaction rates for this specific example to
compensate for the fact that solving differential equations in MatLab
is faster than executing a concurrent model.

Finally, our high level notation is an abstraction for both SPiM and
chemical reactions. %We also show how to translate an arbitrary higher
%order chemical reaction, showing the generality of the approach.  
Our high level notation corresponds to commonly-occurring reaction
schemes easily identifiable by biologists, and it is an alternative to
the narrative language proposed for SPiM. Our high level notation
represents the G-protein cycle quite concisely, it is accessible to
the average biologist, and it translates naturally into SPiM and
chemical reactions, suggesting that our models could be effective
instruments to assist in biomedical research.

This notation naturally induces a typed modeling language that is
currently being explored, where 
a,b~$\vdash$ bind@r;P reduces to
\textit{complex}(a,b) $\vdash$ P, where \textit{complex} is only
defined for specific species, for example.

{\small
\bibliographystyle{eptcs}
\bibliography{refs}

\begin{thebibliography}{10}
\providecommand{\bibitemstart}[1]{\bibitem{#1}}
\providecommand{\bibitemend}{}
\providecommand{\bibliographystart}{}
\providecommand{\bibliographyend}{}
\providecommand{\url}[1]{\texttt{#1}}
\providecommand{\urlprefix}{Available at }
\providecommand{\bibinfo}[2]{#2}
\bibliographystart

\bibitemstart{G-Protein-SPiM}
\bibinfo{author}{Yifei Bao}, \bibinfo{author}{Tommy White},
  \bibinfo{author}{Joseph Glavy} \& \bibinfo{author}{Adriana Compagnoni}
  (\bibinfo{year}{2010}): \emph{\bibinfo{title}{Application of SPiM to Process
  Modeling for the Activation Cycle of G-proteins by G-protein-coupled
  Receptors}}.
\newblock \bibinfo{type}{Technical Report} \bibinfo{number}{CS-2010-1},
  \bibinfo{institution}{Stevens Institute of Technology}.
\bibitemend

\bibitemstart{Bockaert03}
\bibinfo{author}{Joël Bockaert}, \bibinfo{author}{Philippe Marin},
  \bibinfo{author}{Aline Dumuis} \& \bibinfo{author}{Laurent Fagni}
  (\bibinfo{year}{2003}): \emph{\bibinfo{title}{The `magic tail' of G
  Protein-Coupled Receptors: An Anchorage for Functional Protein Networks}}.
\newblock {\sl \bibinfo{journal}{FEBS Letters}}
  \bibinfo{volume}{546}(\bibinfo{number}{1}), pp. \bibinfo{pages}{65--72}.
\newblock \bibinfo{note}{Signal Transduction Special Issue}.
\bibitemend

\bibitemstart{Cardelli08}
\bibinfo{author}{Luca Cardelli} (\bibinfo{year}{2008}):
  \emph{\bibinfo{title}{From Processes to ODEs by Chemistry}}.
\newblock In: {\sl \bibinfo{booktitle}{IFIP TCS}}, pp.
  \bibinfo{pages}{261--281}.
\bibitemend

\bibitemstart{Cardelli07}
\bibinfo{author}{Luca Cardelli} (\bibinfo{year}{2008}):
  \emph{\bibinfo{title}{On Process Rate Semantics}}.
\newblock {\sl \bibinfo{journal}{Theor. Comput. Sci.}}
  \bibinfo{volume}{391}(\bibinfo{number}{3}), pp. \bibinfo{pages}{190--215}.
\bibitemend

\bibitemstart{Artificial}
\bibinfo{author}{Luca Cardelli} (\bibinfo{year}{2009}):
  \emph{\bibinfo{title}{Artificial Biochemistry}}.
\newblock In: \bibinfo{editor}{Anne Condon}, \bibinfo{editor}{David Harel},
  \bibinfo{editor}{Joost~N. Kok}, \bibinfo{editor}{Arto Salomaa} \&
  \bibinfo{editor}{Erik Winfree}, editors: {\sl \bibinfo{booktitle}{Algorithmic
  Bioprocesses}}, \bibinfo{series}{Natural Computing Series},
  \bibinfo{publisher}{Springer Berlin Heidelberg}, pp.
  \bibinfo{pages}{429--462}.
\bibitemend

\bibitemstart{CardelliActin09}
\bibinfo{author}{Luca Cardelli}, \bibinfo{author}{Emmanuelle Caron},
  \bibinfo{author}{Philippa Gardner}, \bibinfo{author}{Ozan Kahramanogullari}
  \& \bibinfo{author}{Andrew Phillips} (\bibinfo{year}{2009}):
  \emph{\bibinfo{title}{A Process Model of Actin Polymerisation}}.
\newblock {\sl \bibinfo{journal}{Electr. Notes Theor. Comput. Sci.}}
  \bibinfo{volume}{229}(\bibinfo{number}{1}), pp. \bibinfo{pages}{127--144}.
\bibitemend

\bibitemstart{CardelliGTP09}
\bibinfo{author}{Luca Cardelli}, \bibinfo{author}{Emmanuelle Caron},
  \bibinfo{author}{Philippa Gardner}, \bibinfo{author}{Ozan Kahramanogullari}
  \& \bibinfo{author}{Andrew Phillips} (\bibinfo{year}{2009}):
  \emph{\bibinfo{title}{A Process Model of Rho GTP-binding Proteins}}.
\newblock {\sl \bibinfo{journal}{Theor. Comput. Sci.}}
  \bibinfo{volume}{410}(\bibinfo{number}{33-34}), pp.
  \bibinfo{pages}{3166--3185}.
\bibitemend

\bibitemstart{tool09}
\bibinfo{author}{Federica Ciocchetta}, \bibinfo{author}{Adam Duguid},
  \bibinfo{author}{Stephen Gilmore}, \bibinfo{author}{Maria~Luisa Guerriero} \&
  \bibinfo{author}{Jane Hillston} (\bibinfo{year}{2009}):
  \emph{\bibinfo{title}{The Bio-PEPA Tool Suite}}.
\newblock {\sl \bibinfo{journal}{Quantitative Evaluation of Systems,
  International Conference on}} \bibinfo{volume}{0}, pp.
  \bibinfo{pages}{309--310}.
\bibitemend

\bibitemstart{Biopepa09}
\bibinfo{author}{Federica Ciocchetta} \& \bibinfo{author}{Jane Hillston}
  (\bibinfo{year}{2009}): \emph{\bibinfo{title}{Bio-PEPA: A framework for the
  Modelling and Analysis of Biological Systems}}.
\newblock {\sl \bibinfo{journal}{Theoretical Computer Science}}
  \bibinfo{volume}{410}(\bibinfo{number}{33-34}), pp.
  \bibinfo{pages}{3065--3084}.
\bibitemend

\bibitemstart{DV09}
\bibinfo{author}{J\'{e}r\^{o}me Feret}, \bibinfo{author}{Vincent Danos},
  \bibinfo{author}{Jean Krivine}, \bibinfo{author}{Russ Harmer} \&
  \bibinfo{author}{Walter Fontana} (\bibinfo{year}{2009}):
  \emph{\bibinfo{title}{{Internal Coarse-graining of Molecular Systems}}}.
\newblock {\sl \bibinfo{journal}{Proceedings of the National Academy of
  Sciences}} \bibinfo{volume}{106}(\bibinfo{number}{16}), pp.
  \bibinfo{pages}{6453--6458}.
\bibitemend

\bibitemstart{Maria07}
\bibinfo{author}{Maria~Luisa Guerriero}, \bibinfo{author}{John~K. Heath} \&
  \bibinfo{author}{Corrado Priami} (\bibinfo{year}{2007}):
  \emph{\bibinfo{title}{An Automated Translation from a Narrative Language for
  Biological Modelling into Process Algebra}}.
\newblock In: {\sl \bibinfo{booktitle}{CMSB}}, pp. \bibinfo{pages}{136--151}.
\bibitemend

\bibitemstart{Hao03}
\bibinfo{author}{Nan Hao}, \bibinfo{author}{Necmettin Yildirim},
  \bibinfo{author}{Yuqi Wang}, \bibinfo{author}{Timothy~C. Elston} \&
  \bibinfo{author}{Henrik~G. Dohlman} (\bibinfo{year}{2003}):
  \emph{\bibinfo{title}{Regulators of G Protein Signaling and Transient
  Activation of Signaling: Experimental and computational analysis reveals
  negative and positive feedback controls on g protein activity}}.
\newblock {\sl \bibinfo{journal}{J. Biol. Chem.}}
  \bibinfo{volume}{278}(\bibinfo{number}{47}), pp.
  \bibinfo{pages}{46506--46515}.
\bibitemend

\bibitemstart{Pepa96}
\bibinfo{author}{Jane Hillston} (\bibinfo{year}{1996}): \emph{\bibinfo{title}{A
  Compositional Approach to Performance Modelling}}.
\newblock \bibinfo{publisher}{Cambridge University Press},
  \bibinfo{address}{New York, NY, USA}.
\bibitemend

\bibitemstart{Howard01}
\bibinfo{author}{Andrew~D. Howard}, \bibinfo{author}{George McAllister},
  \bibinfo{author}{Scott~D. Feighner}, \bibinfo{author}{Qingyun Liu},
  \bibinfo{author}{Ravi~P. Nargund} \& \bibinfo{author}{Lex H. T.~Van der
  Ploeg~an} (\bibinfo{year}{2001}): \emph{\bibinfo{title}{Orphan
  G-protein-coupled Receptors and Natural Ligand Discovery}}.
\newblock {\sl \bibinfo{journal}{Trends in Pharmacological Sciences}}
  \bibinfo{volume}{22}(\bibinfo{number}{3}), pp. \bibinfo{pages}{132--140}.
\bibitemend

\bibitemstart{Ozan09}
\bibinfo{author}{Ozan Kahramanogullari}, \bibinfo{author}{Luca Cardelli} \&
  \bibinfo{author}{Emmanuelle Caron} (\bibinfo{year}{2009}):
  \emph{\bibinfo{title}{An Intuitive Automated Modelling Interface for Systems
  Biology}}.
\newblock {\sl \bibinfo{journal}{CoRR}} \bibinfo{volume}{abs/0911.2327}.
\bibitemend

\bibitemstart{Kobilka06}
\bibinfo{author}{Brian~K. Kobilka} (\bibinfo{year}{2007}):
  \emph{\bibinfo{title}{G Protein Coupled Receptor Structure and Activation.}}
\newblock {\sl \bibinfo{journal}{Biochim Biophys Acta}}
  \bibinfo{volume}{1768}(\bibinfo{number}{4}), pp. \bibinfo{pages}{794--807}.
\bibitemend

\bibitemstart{Lousi06}
\bibinfo{author}{Louis~M. Luttrell} (\bibinfo{year}{2006}):
  \emph{\bibinfo{title}{Transmembrane Signaling Protocols}}, chapter
  \bibinfo{chapter}{Transmembrane Signaling by G Protein-Coupled Receptors},
  pp. \bibinfo{pages}{3--49}.
\newblock \bibinfo{publisher}{Humana Press}.
\bibitemend

\bibitemstart{Oldham96}
\bibinfo{author}{William~M. Oldham} \& \bibinfo{author}{Heidi~E. Hamm}
  (\bibinfo{year}{2008}): \emph{\bibinfo{title}{Heterotrimeric G protein
  Activation by G-protein-coupled Receptors}}.
\newblock {\sl \bibinfo{journal}{Nat Rev Mol Cell Biol}}
  \bibinfo{volume}{9}(\bibinfo{number}{1}), pp. \bibinfo{pages}{60--71}.
\bibitemend

\bibitemstart{Priami09}
\bibinfo{author}{Alida Palmisano}, \bibinfo{author}{Ivan Mura} \&
  \bibinfo{author}{Corrado Priami} (\bibinfo{year}{2009}):
  \emph{\bibinfo{title}{From ODES to Language-based, Executable Models of
  Biological Systems}}.
\newblock {\sl \bibinfo{journal}{Pacific Symposium on Biocomputing}}
  \bibinfo{volume}{14}(\bibinfo{number}{12}), pp. \bibinfo{pages}{239--250}.
\bibitemend

\bibitemstart{mor05}
\bibinfo{author}{Mor Peleg}, \bibinfo{author}{Daniel Rubin} \&
  \bibinfo{author}{Russ~B. Altman} (\bibinfo{year}{2005}):
  \emph{\bibinfo{title}{{Using Petri Net Tools to Study Properties and Dynamics
  of Biological Systems}}}.
\newblock {\sl \bibinfo{journal}{Journal of the American Medical Informatics
  Association}} \bibinfo{volume}{12}(\bibinfo{number}{2}), pp.
  \bibinfo{pages}{181--199}.
\bibitemend

\bibitemstart{Phillips09}
\bibinfo{author}{Andrew Phillips} (\bibinfo{year}{2009}):
  \emph{\bibinfo{title}{Symbolic Systems Biology: Theory and Methods}}, chapter
  \bibinfo{chapter}{A Visual Process Calculus for Biology}.
\newblock \bibinfo{publisher}{Jones and Bartlett}.
\newblock \bibinfo{note}{In Press}.
\bibitemend

\bibitemstart{Phillips07}
\bibinfo{author}{Andrew Phillips} \& \bibinfo{author}{Luca Cardelli}
  (\bibinfo{year}{2007}): \emph{\bibinfo{title}{Efficient, Correct Simulation
  of Biological Processes in the Stochastic Pi-calculus}}.
\newblock In: {\sl \bibinfo{booktitle}{Computational Methods in Systems
  Biology}}, {\sl \bibinfo{series}{LNCS}} \bibinfo{volume}{4695},
  \bibinfo{publisher}{Springer}, pp. \bibinfo{pages}{184--199}.
\bibitemend

\bibitemstart{Pierce02}
\bibinfo{author}{Kristen~L. Pierce}, \bibinfo{author}{Richard~T. Premont} \&
  \bibinfo{author}{Robert~J. Lefkowitz} (\bibinfo{year}{2002}):
  \emph{\bibinfo{title}{Seven-transmembrane Receptors}}.
\newblock {\sl \bibinfo{journal}{Nat Rev Mol Cell Biol}}
  \bibinfo{volume}{3}(\bibinfo{number}{9}), pp. \bibinfo{pages}{639--650}.
\bibitemend

\bibitemstart{JW03}
\bibinfo{author}{John~W. Pinney}, \bibinfo{author}{David~R. Westhead} \&
  \bibinfo{author}{Glenn~A. McConkey} (\bibinfo{year}{2003}):
  \emph{\bibinfo{title}{Petri Net representations in Systems Biology.}}
\newblock {\sl \bibinfo{journal}{Biochem. Soc. Trans.}}
  \bibinfo{volume}{31}(\bibinfo{number}{6}), pp. \bibinfo{pages}{1513--1515}.
\bibitemend

\bibitemstart{Priami01}
\bibinfo{author}{Corrado Priami}, \bibinfo{author}{Aviv Regev},
  \bibinfo{author}{Ehud~Y. Shapiro} \& \bibinfo{author}{William Silverman}
  (\bibinfo{year}{2001}): \emph{\bibinfo{title}{Application of A Stochastic
  Name-passing Calculus to Representation and Simulation of Molecular
  Processes}}.
\newblock {\sl \bibinfo{journal}{Inf. Process. Lett.}}
  \bibinfo{volume}{80}(\bibinfo{number}{1}), pp. \bibinfo{pages}{25--31}.
\bibitemend

\bibitemstart{Robishaw04}
\bibinfo{author}{Janet~D. Robishaw} \& \bibinfo{author}{Catherine~H. Berlot}
  (\bibinfo{year}{2004}): \emph{\bibinfo{title}{Translating G protein Subunit
  Diversity Into Functional Specificity}}.
\newblock {\sl \bibinfo{journal}{Current Opinion in Cell Biology}}
  \bibinfo{volume}{16}(\bibinfo{number}{2}), pp. \bibinfo{pages}{206--209}.
\bibitemend

\bibitemstart{vilar02}
\bibinfo{author}{Jos\'{e} M.~G. Vilar}, \bibinfo{author}{Hao~Yuan Kueh},
  \bibinfo{author}{Naama Barkai} \& \bibinfo{author}{Stanislas Leibler}
  (\bibinfo{year}{2002}): \emph{\bibinfo{title}{Mechanisms of Noise-resistance
  in Genetic Oscillators}}.
\newblock {\sl \bibinfo{journal}{PNAS}}
  \bibinfo{volume}{99}(\bibinfo{number}{9}), pp. \bibinfo{pages}{5988--5992}.
\bibitemend

\bibitemstart{Yi2003}
\bibinfo{author}{Tau~M. Yi}, \bibinfo{author}{Hiroaki Kitano} \&
  \bibinfo{author}{Melvin~I. Simon} (\bibinfo{year}{2003}):
  \emph{\bibinfo{title}{A Quantitative Characterization of The Yeast
  Heterotrimeric G Protein Cycle.}}
\newblock {\sl \bibinfo{journal}{Proc Natl Acad Sci U S A}}
  \bibinfo{volume}{100}(\bibinfo{number}{19}), pp. \bibinfo{pages}{10764--9}.
\bibitemend

\bibliographyend
\end{thebibliography}
}
\end{document}